%% file: cl2.tex
\documentstyle{mn}
\input psfig.sty
\input mn_junk.tex

\def\hmpc{{\rm h}$^{-1}~{\rm Mpc}\,$}
\title[Clustering of Galaxy Clusters]
{Clustering of Galaxy Clusters in CDM Universes.}
\author[J. M. Colberg et al.]
{J.~M.~Colberg,$^{1}$ S.~D.~M.~White,$^1$ N.~Yoshida,$^1$ T.~J.~MacFarland,$^{1,2}$ A.~Jenkins,$^3$\\
\newauthor C.~S.~Frenk,$^3$ F.~R.~Pearce,$^3$  A.~E.~Evrard,$^4$ H.~M.~P.~Couchman,$^5$\\
\newauthor G.~Efstathiou,$^6$ J.~A.~Peacock$^7$ and P.A.~Thomas.$^8$ (The Virgo Consortium)\\
$^1$ Max-Planck Inst. for Astrophysics, Garching, Munich, D-85740, Germany.\\
$^2$ Now at 105 Lexington Avenue, Apt. 6F,New York, NY 10016\\
$^3$ Dept Physics, South Road, Durham, DH1 3LE.\\
$^4$ Dept Physics, University of Michigan, Ann Arbor, MI-48109-1120.\\
$^5$ Department of Physics and Astronomy, McMaster University, Hamilton,
Ontario, L8S 4M1, Canada\\
$^6$ Institute of Astronomy, Madingley Road, Cambridge CB3 OHA\\
$^7$ Royal Observatory, Institute of Astronomy, Edinburgh, EH9 3HJ\\
$^8$ Astronomy Centre,CPES, University of Sussex, Falmer, Brighton, BN1 9QH.}
\date{10/7/00}
\pagerange{\pageref{firstpage}--\pageref{lastpage}}
\pubyear{2000}

\begin{document}
\label{firstpage}

\maketitle

\begin{abstract}

We use very large cosmological N--body simulations to obtain accurate
predictions for the two-point correlations and power spectra of
mass-limited samples of galaxy clusters.  We consider two currently
popular cold dark matter (CDM) cosmogonies, a critical density model
($\tau$CDM) and a flat low density model with a cosmological constant
($\Lambda$CDM). Our simulations each use $10^9$ particles to follow
the mass distribution within cubes of side $2h^{-1}$Gpc ($\tau$CDM)
and $3h^{-1}$Gpc ($\Lambda$CDM) with a force resolution better than
$10^{-4}$ of the cube side. We investigate how the predicted cluster
correlations increase for samples of increasing mass and decreasing
abundance. Very similar behaviour is found in the two cases. The
correlation length increases from $r_0=12$ -- 13$h^{-1}$Mpc for
samples with mean separation $d_{\rm c}=30h^{-1}$Mpc to $r_0=22$ --
27$h^{-1}$Mpc for samples with $d_{\rm c}=100h^{-1}$Mpc. The lower
value here corresponds to $\tau$CDM and the upper to $\Lambda$CDM. The
power spectra of these cluster samples are accurately parallel to
those of the mass over more than a decade in scale. Both correlation
lengths and power spectrum biases can be predicted to better than 10\%
using the simple model of Sheth, Mo \& Tormen (2000).  This prediction
requires only the linear mass power spectrum and has no adjustable
parameters. We compare our predictions with published results for the
APM cluster sample.  The observed variation of correlation length with
richness agrees well with the models, particularly for
$\Lambda$CDM. The observed power spectrum (for a cluster sample of
mean separation $d_{\rm c}=31h^{-1}$Mpc) lies significantly above the
predictions of both models.
\end{abstract}

\begin{keywords}
cosmology:theory - dark matter - gravitation - galaxy clusters--simulations
\end{keywords}

\section{Introduction}
The last two decades have established cosmological N-body simulations
as the principal tool for studying the evolution of large-scale
structure. The earliest systematic studies used $10^3$ to $2\times
10^4$ particles to follow evolution from white noise or other
similarly {\it ad hoc} initial conditions (Gott, Turner \& Aarseth
1979; Efstathiou \& Eastwood 1981). They showed that nonlinear growth
could produce a power law autocorrelation function similar to that
measured for galaxies. Soon thereafter, the suggestion that the dark
matter might be a weakly interacting massive particle led to the first
simulations from initial conditions based on a detailed treatment of
the physics of earlier evolution. These represented the dark matter
distribution within cubic regions with periodic boundary conditions
using only $3\times 10^4$ particles. They were nevertheless able to
show that, for adiabatic fluctuations produced during inflation, a
neutrino-dominated universe is not viable (White, Frenk \& Davis 1983)
while a cold dark matter (CDM) dominated universe is much more
promising (Davis et al. 1985).

Since this early work, many groups have used their own simulations 
to compare predictions for large-scale structure 
with the wealth of data coming from new 
observational surveys. As algorithms and computers have improved, 
the number of particles treated in high resolution simulations has 
increased. Thus, White et al. (1987a,b) could already use $2.6\times 
10^5$ particles to study CDM universes, while Warren et al. (1992), 
Gelb \& Bertschinger (1994), Jenkins et al. (1998) and Governato et 
al. (1999) studied large-scale structure using high 
resolution simulations with $1\times 10^6$, $3\times 10^6$, $1.7\times
10^7$ and $4.7\times 10^7$ particles, respectively. More particles are
better for two reasons. One can choose to have better mass resolution 
so that the internal properties of each structure are better defined, 
and one can simulate larger volumes so that more structures are included
and the statistical distribution of their properties is better defined.
Here, we report results for the spatial distribution of galaxy
clusters from simulations using $1\times 10^9$ particles. The volumes
simulated are much larger than any attempted previously and are large 
compared even to the biggest currently planned observational surveys. 
As a result our theoretical predictions have high precision. 

The two--point correlation function of galaxy clusters has been
controversial for decades. The early work of Hauser \&
Peebles (1973) showed that rich galaxy clusters are
more strongly clustered than galaxies, and estimates of the
autocorrelation function of Abell clusters by Bahcall \& Soneira
(1983) and Klypin \& Kopylov (1983) agreed on a power-law form 
which parallels the galaxy autocorrelation function but with substantially
greater amplitude. Subsequent work has failed to agree on the strength
of this enhancement and on its dependence on the properties which
define the cluster sample. Thus, Sutherland (1988) argued that much
of the apparent clustering in the original samples was induced 
artificially by Abell's criteria for defining clusters. This
conclusion has been supported by some subsequent studies (e.g. Croft
et al. 1997 and references therein) and disputed by others (e.g.
Olivier et al. 1993 and references therein). 

Although all authors agree that richer clusters are more strongly
clustered, the strength of this trend is also disputed. Bahcall and
co-workers (e.g. Bahcall \& Cen 1992, Bahcall \& West 1992) have 
argued that the correlation length, $r_0$, defined via
$\xi_{cl}(r_0)=1$, scales linearly with intercluster separation,
$d_{\rm c}$,
\begin{equation}
r_0 = 0.4\,d_{\rm c} = 0.4\, n_{\rm c}^{-1/3}\,,
\label{eq:linear}
\end{equation} 
where $n_{\rm c}$ is the number density of clusters above the chosen
richness threshold.  This scaling might be expected in a fractal model
of large-scale structure (Szalay \& Schramm 1985) and appeared
consistent with early measurements for Abell clusters (e.g.\ Bahcall
\& Soneira 1983). Other work has suggested that this apparent scaling
reflects incompleteness in the Abell samples (e.g.\ Efstathiou et al.\
1992, Peacock \& West 1992). The more objectively defined APM cluster
sample appears to show a significantly weaker trend of clustering
strength with richness (Efstathiou 1996; Croft et al. 1997).
\begin{table*}
\centering
\begin{minipage}{140mm}
\caption{Parameters of the Hubble Volume simulations}
\begin{tabular}{@{}lrrrrrc@{}}
Model &        $\Omega$ & $\Lambda$ & $h$ & $\Gamma$ & $\sigma_8$ & $L_{\rm box}$\\
$\tau$CDM    &   1.0    &    0.0    & 0.5 &  0.21    &   0.6    & 2000 Mpc/$h$ \\
$\Lambda$CDM &   0.3    &    0.7    & 0.7 &  0.17    &   0.9    & 3000 Mpc/$h$\\
\end{tabular}\label{ttable}
\end{minipage}
\end{table*} 

Quite surprisingly, both camps have used N--body simulations of standard
CDM cosmogonies to support their views. Bahcall \& Cen (1992) found $r_0$
to increase roughly in proportion to $d_{\rm c}$ for their simulated
clusters, while Croft \& Efstathiou (1994) found a weaker dependence.  The
latter authors found their cluster correlation function to be {\it
insensitive} to the cosmic matter density, $\Omega$, and to depend weakly
on the normalization of the power spectrum, $\sigma_8$, but strongly on its
shape. (Here, $\sigma_8$ denotes linearly extrapolated present-day rms mass
fluctuation in spherical top hat spheres of radius 8\hmpc.)  Similar
conclusions were reached by Eke \etal (1996) who studied systematics in
simulated cluster correlations, in particular the influence of the
definition of a cluster. They argued that the different scalings of $r_0$
with $d_{\rm c}$ seen in previous N-body simulations stemmed primarily from
the use of different algorithms to identify clusters in the simulations.
All this work suffered from the relatively small volumes simulated, which
limited the statistical accuracy of the correlation estimates, especially
for rare and massive clusters. A substantial improvement came with the work
of Governato et al. (1999) who used more particles and treated
significantly larger volumes. The simulations presented below provide a
further major improvement by using 20 times as many particles and
increasing the volumes treated by about two orders of magnitude.

The results we present below are in general agreement with those of
Governato et al. (1999), but our work achieves substantially higher
statistical precision. We show very clearly that the strength of
cluster correlations is predicted to increase significantly with
cluster richness in currently popular CDM cosmogonies.  Furthermore,
these correlations can be predicted remarkably accurately (and with
{\it no} free parameters) by the recent analytic model of Sheth, Mo \& Tormen
(1996). In this model, which refines that of Mo \& White
(1996),  the two--point correlation function of dark
haloes is proportional to that of the dark matter, the ratio of the
two depending on halo mass and on the {\it linear} power spectrum of
mass density fluctuations (see below). Mo \& White tested their original 
model on a set of scale-free N--body simulations, finding good qualitative
agreement. For CDM models, Sheth, Mo \& Tormen found the quantitative
prediction both of halo mass functions and of halo correlations to be
improved substantially by generalising the Mo \& White approach to
ellipsoidal (rather than spherical) collapse. Our results here reach
higher precision and extend these tests to rarer objects;
a preliminary account was published in Colberg et al. (1998), which is 
superseded by the current paper.

The second order statistics of the spatial distribution
of clusters can, of course, be analysed using power spectra rather
than correlation functions. Such an approach is particularly
advantageous for analysing fluctuations on large spatial scales.
Recent observational estimates of the cluster power spectrum
have been given by Borgani et al. (1997) and
Retzlaff et al. (1998), and by Tadros, Efstathiou \& Dalton (1998) for
the Abell--ACO 
and APM clusters, respectively. For both samples, there is an
indication of a peak in the power spectrum at a wavenumber of
$k\sim 0.03\,h\,\mbox{Mpc}^{-1}$. This is roughly coincident with
the scale where a peak is expected for
currently popular CDM models. The simulation data we present below
verify that the cluster power spectrum should indeed parallel that 
of the mass on these scales.

In the following Section we briefly discuss the Hubble Volume 
simulations and the way we have defined cluster samples
within them. In Section~\ref{tpcf}, we present two--point correlations 
for these samples and compare them with the analytic model. In
Section~\ref{power}, we present power spectra for samples 
constructed to correspond directly to the APM cluster survey; an
interesting result is that the observations and predictions are
significantly discrepant for the current ``best buy'' cosmogony.
We conclude with a summary of our main results.
\begin{figure*}
\centerline{\psfig{figure=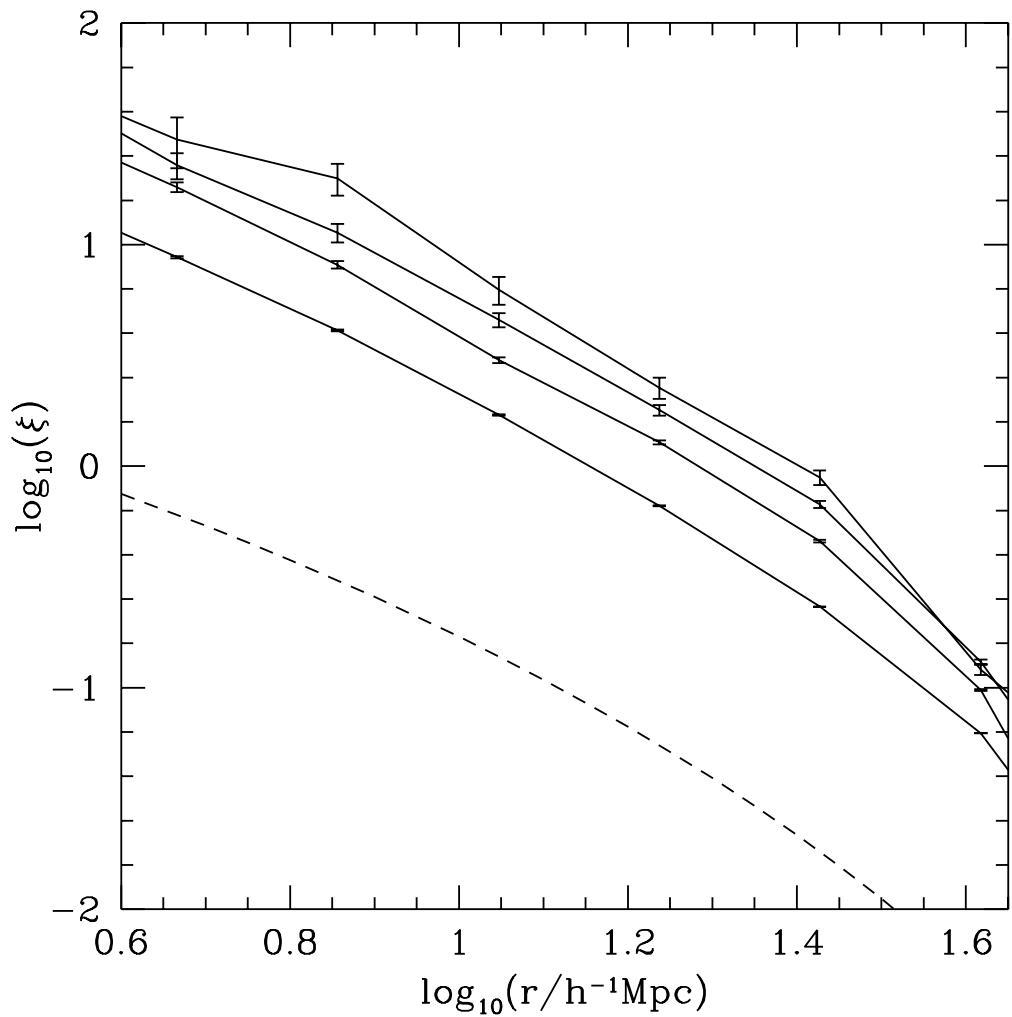,width=400pt,height=400pt}}
\caption{Two--point correlation functions of the $\tau$CDM 
model for the $d_{\rm c}=40$, 70, 100 and 130 $h^{-1}$Mpc samples (solid
lines, from bottom to top). The plotted 1$\sigma$ errorbars are
derived from the number of pairs in each bin. The dashed line is the
two--point correlation of the dark matter.}\label{fig:tpcf_tCDM}
\end{figure*}
\begin{figure*}
\centerline{\psfig{figure=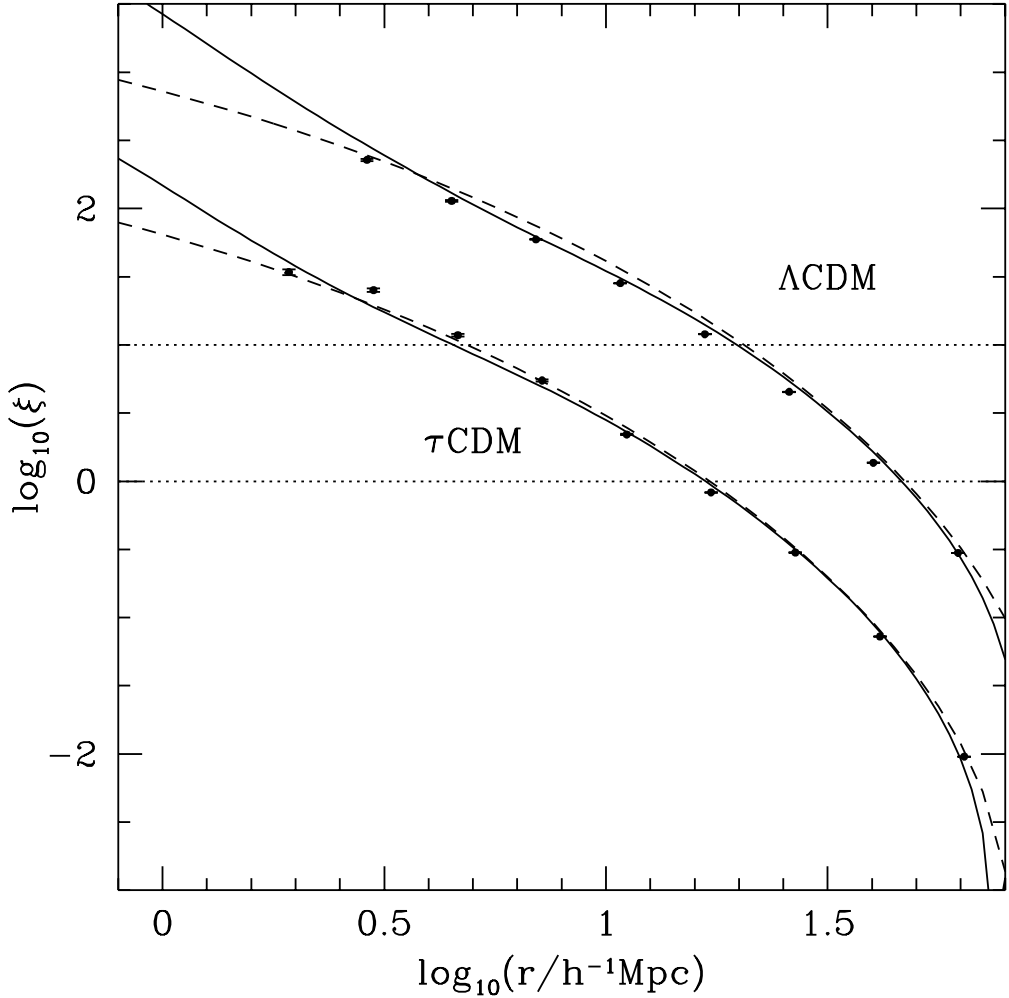,width=400pt,height=400pt}}
\caption{The two--point correlation functions of the $\tau$CDM (lower
plots) and $\Lambda$CDM (upper plots) models for 
$d_c=50 h^{-1}$Mpc. This figure compares results from the simulations
(dots with errorbars) with the linear (dashed line) and nonlinear 
(solid line) predictions from eqn~\ref{eq:mo} with the SMT prediction
for $b$. For the $\Lambda$CDM
model all quantities have been shifted upwards by one order of magnitude.
1$\sigma$ errorbars are plotted, as in fig.\ \ref{fig:tpcf_tCDM}.}
\label{fig:tpcf_single}
\end{figure*}

\section{Clusters in the Hubble Volume Simulations}\label{clusters}

The two simulations analysed in this paper were carried out in 1997
and 1998 on 512 processors of the CRAY T3E at the Garching Computer
Centre of the Max Planck Society. They used a specially stripped down
version of parallel Hydra, the workhorse code of the Virgo
Supercomputing Consortium. Details may be found in MacFarland et
al. (1998).  This code maximises the efficiency of memory use on the
machine and allowed the trajectories of $10^9$ particles to be
integrated accurately, with a gravitational force resolution of about
$10^{-4}$ of the side of the computational volume. Each simulation
used about 50,000 processor hours of CPU time. The two cases studied
were a $(2000 h^{-1}{\rm Mpc})^3$ volume of a $\tau$CDM universe and a
$(3000 h^{-1}{\rm Mpc})^3$ volume of a $\Lambda$CDM universe. In both
cases the mass of a single particle is $2\times
10^{12}h^{-1}M_{\odot}$ and the simulation is normalized to yield the
observed abundance of rich clusters at $z=0$ (White, Efstathiou \&
Frenk 1993; Eke, Cole \& Frenk 1996). These normalisations are also
consistent with the level of fluctuations measured by COBE. The
parameters of the simulations are summarized in table
\ref{ttable}. (Here, $\Gamma$ denotes the spectral shape parameter;
c.f. Efstathiou, Bond \& White 1992.)

The Hubble Volume simulations are essentially larger realisations of
two of the cosmological models previously simulated by Jenkins \etal\
(1998).  We have checked that basic properties of our new simulations,
such as the mass power spectrum and the velocity field, are consistent
with expectations based on our smaller simulations. High order
clustering statistics in the $\tau$CDM Hubble Volume simulation have
been extensively studied by Szapudi \etal\ (2000) and Colombi
\etal\ (2000). Both Hubble Volume simulations were
used by Jenkins \etal\ (2000) in a study of the mass function of dark
matter halos.  The mass functions from the Hubble simulations are
consistent with those from smaller simulations in the regions of
overlap.

Clusters of galaxies were identified in these simulations using a
spherical overdensity (SO) group finder (Lacey \& Cole 1994).  This
defines the cluster boundary as the sphere within which the mean
density is 180 and 324 times the critical value in the $\tau$CDM and
$\Lambda$CDM cases respectively. The lowest mass  clusters
considered in our analysis have 75 and 39 particles respectively in
the $\tau$CDM and $\Lambda$CDM models.  We have checked that our
results in the form we present below are insensitive to this
choice. For example, we obtain almost identical results if clusters
are defined using a friends-of-friends algorithm (Davis et al. 1985)
with linking lengths of 0.2 and 0.164 in each model (which produces
clusters with at least 86 and 44 particles respectively.)  The choice
of grouping algorithm and associated parameters affects the masses
assigned to clusters in a systematic way, but has no significant
systematic effect on their positions or on their ranking in mass.  

We construct a series of mass-limited cluster catalogues and
characterise each one by the mean separation $d_{\rm c}$ of the
clusters it contains. The advantage of this parameterisation is that
it allows a precise comparison with observed richness-limited samples
and with analytic models without any need to ensure that the mass
definitions in the three cases correspond exactly. In
the current paper, we consider only the clustering of clusters at
$z=0$. The excellent fits we find to the analytic theory also hold at
other redshifts, so in practice the analytic formulae can be used to
describe superclustering at any redshift.
\begin{figure*}
\centerline{\psfig{figure=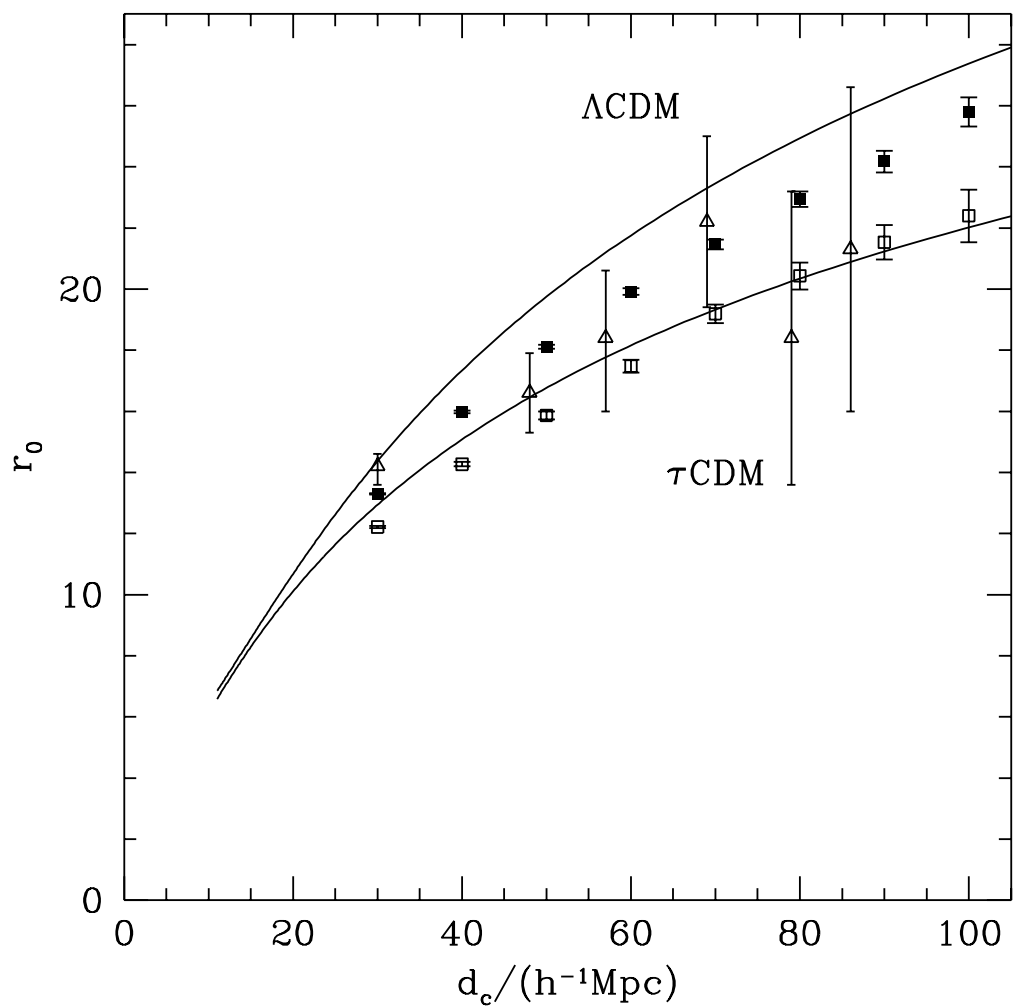,width=400pt,height=400pt}}
\caption{Correlation length, $r_0$, as a function of mean intercluster
separation, $d_c$,  for the $\tau$CDM (open squares) and
$\Lambda$CDM (filled squares) simulations. The predictions of 
the SMT model are shown as solid lines. Also shown are  
data from the APM cluster catalogue (open triangles), taken from 
Croft et al.\ (1997).}\label{fig:tpcf_r0}
\end{figure*}

\begin{figure*}
\centerline{\psfig{figure=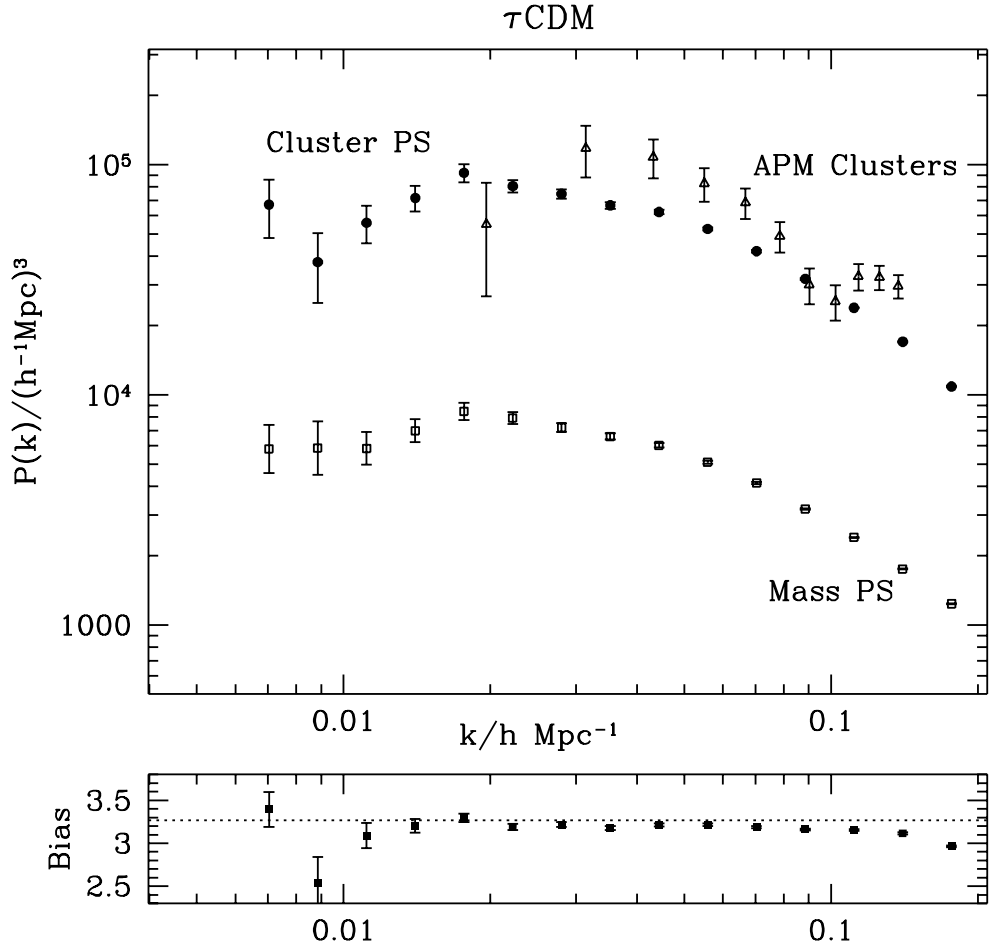,width=400pt,height=400pt}}
\caption{The upper panel shows power spectra for galaxy clusters with 
$d_{\rm c}=30.9\,h^{-1}$\,Mpc from the $\tau$CDM simulation
(filled dots), for the APM cluster sample (triangles; taken from
Tadros et al.\ 1998), and for the dark matter in the simulation (open
squares). The lower panel gives a bias factor defined as the square root
of the ratio of the cluster and dark matter power spectra. The
horizontal dotted line is the value of this bias predicted by the 
SMT model.}\label{fig:power_tcdm}
\end{figure*}

\begin{figure*}
\centerline{\psfig{figure=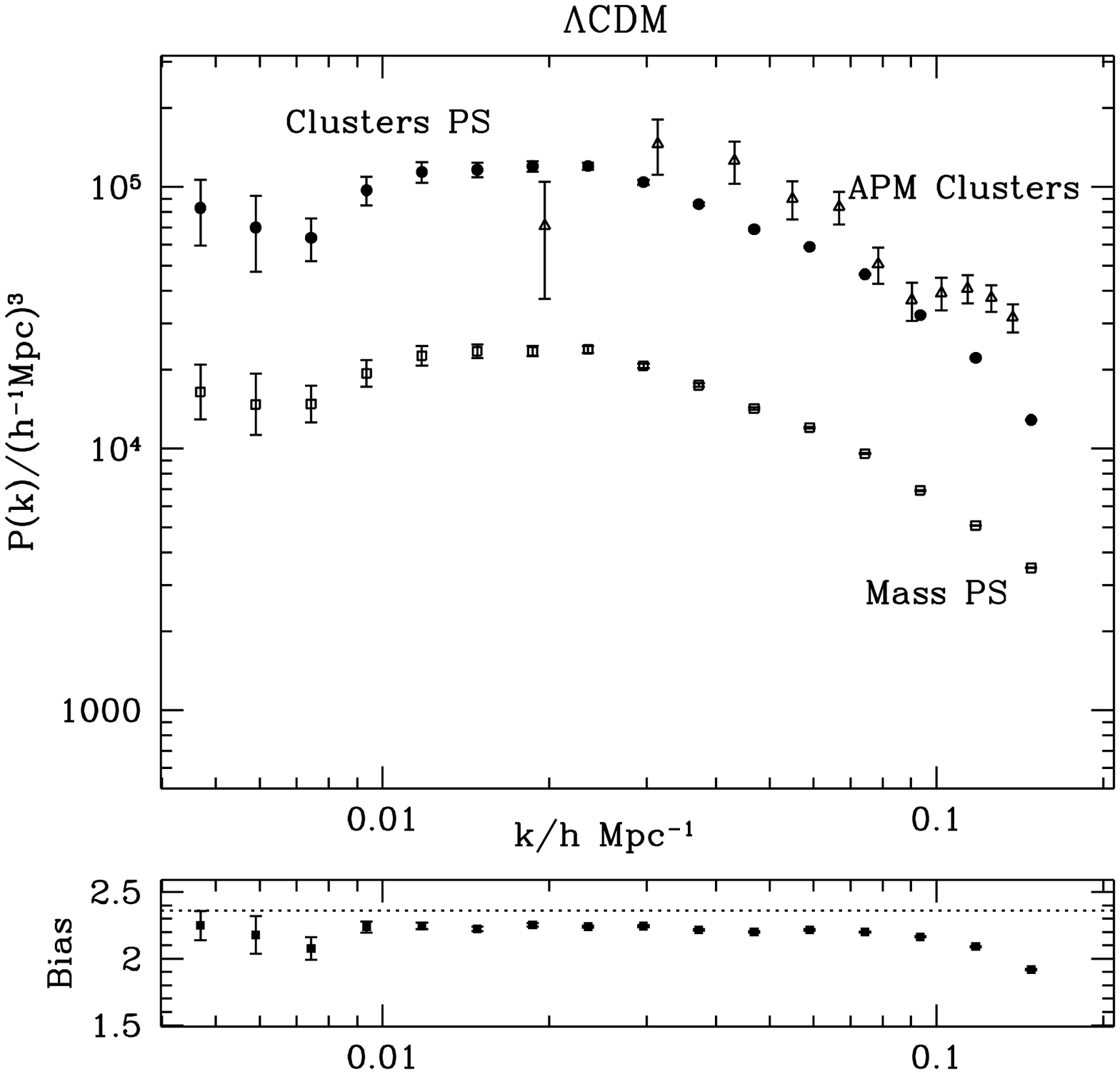,width=400pt,height=400pt}}
\caption{As in fig.\ \ref{fig:power_tcdm} but for the $\Lambda$CDM model.
For this model the SMT prediction of the bias factor is high but only
by 6\%. }
\label{fig:power_lcdm}
\end{figure*}

\section{Two--point Correlation Functions} \label{tpcf}

\subsection{The Analytic Model}
Starting from a ``Press--Schechter'' (1974) argument similar to those
in Bond et al. (1991) and Lacey \& Cole (1993), Mo \& White (1996)
developed an analytic theory for the spatial clustering of dark
haloes in hierarchical clustering models such as the many variants
of CDM.  They find that the two--point correlation
function of dark matter haloes of mass $M$ may be approximately related
to that of the mass by
\begin{equation}
\xi_h(r; M) = b^2(M)\,\xi (r)\,,
\label{eq:mo}
\end{equation}
where
\begin{equation} 
b(M) = 1 + \frac{\delta_{\rm c}}{\sigma^2(M)} - \frac{1}{\delta_{\rm c}}\,.
\label{eq:mob}
\end{equation}
Here, $\delta_{\rm c}=1.686$ is the interpolated linear overdensity at
collapse of a spherical perturbation, and $\sigma(M)$ is the {\it
rms} linear fluctuation in overdensity within a sphere which on
 average contains mass $M$. 
 Notice that although $\sigma(M)$ can be calculated directly from
 linear theory, $\xi(r)$ in eqn~\ref{eq:mo} is the full nonlinear
 correlation function of the mass density field. This can be estimated
 from the linear-theory power spectrum
 using, for example, the approximation of Peacock \& Dodds (1996). Thus,
 the nonlinear correlation function of haloes can be predicted
 without the need to carry out an $N$-body simulation. As shown by Cole \&
 Kaiser (1989), eqn~\ref{eq:mob} can be derived by calculating how
 the Press-Schechter mass function responds to small changes
 in the threshold $\delta_c$. 

 It has long been known that the Press-Schechter mass
 function is not a perfect match to the mass functions found in
 simulations (e.g. Efstathiou et al. 1988), and recent work has
 demonstrated that there is a corresponding systematic
 shift in the bias calculated using the Cole-Kaiser argument 
 (Jing 1998; Sheth \& Tormen 1999). Sheth, Mo \& Tormen  (2000; SMT) 
 have shown how the inclusion of a mass-dependent absorbing barrier
 in the excursion set derivation of the mass function (Bond et al. 1991) 
 can model the anisotropic collapse of cosmic structure and
 substantially improve the agreement between analytic theory and
 numerical simulation. Following the logic of Mo \& White's
 extension of the excursion set formalism
 but using this mass-dependent threshold, SMT
 predict halo clustering in good agreement with simulation data.
 For our purposes, the effect of the SMT revision is to predict a
 slightly different $b(M)$ from that in eqn.~(3).
 
A technical problem arises when comparing such analytic formulae with
simulations; it is unclear how to define the boundaries of simulated
clusters so that their mass corresponds to the mass $M$ in
eqn~\ref{eq:mob}. Although this might seem to introduce an additional
degree of freedom, we can eliminate it by using the corresponding
analytic expression for the abundance of clusters 
to convert from sample limiting mass, $M$, to mean cluster
separation $d_{\rm c}$. The predicted correlations can then be
compared to those of a mass-limited sample of simulated clusters with
the same mean separation. This comparison has no adjustable
parameters. Note that for such mass-limited samples the factor $b^2$
in eqn~\ref{eq:mo} is the square of the mean bias obtained by
weighting $b(M)$ by the abundance of clusters of that mass (see, for
example, Baugh et al. 1998 and Governato et al. 1999).

\subsection{Results} \label{results}

Figure \ref{fig:tpcf_tCDM} shows cluster correlation functions
for the $\tau$CDM simulation for mass-limited samples of clusters
with mean separations of 40, 70, 100 and 130 $h^{-1}$Mpc. These
samples contain 125,000, 23,000, 8,000 and 3,600 clusters respectively. We
have computed 1$\sigma$ errors from the numbers of pairs in each
separation bin. Clearly, more massive clusters are more strongly
clustered. Note also the very small error bars on these correlation
estimates which are a consequence of the very large volume of our
simulations.

Figure \ref{fig:tpcf_single} shows the correlation functions of
samples with $d_{\rm c}=50h^{-1}$Mpc from our two simulations,
together with predictions from the SMT model of the last
subsection. The predictions are shown separately for the two cases
where $\xi(r)$ is simply taken as the Fourier transform of the linear
power spectrum, and where it is estimated using the nonlinear model of
Peacock \& Dodds (1996). The correlation functions are very similar in
the two cosmologies, showing that the strength of superclustering is
not a good estimator of $\Omega$ for CDM models normalised to match
the observed abundance of clusters and having a mass correlation
function with a similar shape to the {\it galaxy} correlation function
on large scales. (Note that the curves for $\Lambda$CDM have been
raised by an order of magnitude for clarity.)  The analytic
predictions are in excellent agreement with the numerical results, 
particularly for $\xi_h\sim 1$. Over the relevant range of scales the
linear and nonlinear predictions for $\xi(r)$ are quite close, and
using the nonlinear formula gives at best a marginal improvement in
the fit to the simulation results.

In figure \ref{fig:tpcf_r0} we quantify the increase in clustering
strength with cluster mass by plotting the correlation length, $r_0$,
of our mass-limited cluster samples as a function of their mean
intercluster separation, $d_{\rm c}$. We estimate correlation lengths
from plots similar to those of figure
\ref{fig:tpcf_tCDM} by interpolating between the points on either
side of $\xi_h = 1$. This figure again shows that our simulated
volumes are large enough to estimate correlation lengths with
high accuracy. The values of $r_0$ for $\Lambda$CDM exceed those for
$\tau$CDM by between 10 and 20\%. In both models, the increase in $r_0$
with $d_{\rm c}$ is quite strong, although weaker than the direct 
proportionality suggested by Szalay \& Schramm (1985) and Bahcall \&
Cen (1992). For $\tau$CDM the analytic prediction of $r_0$ is accurate
to within a few percent on all scales; for $\Lambda$CDM it is 
about 10\% high. 

The same general trend of $r_0$ with $d_c$ is also apparent in the
simulations of Governato \etal\ (1999) who considered an
$\Omega_0=0.3$ open CDM model (OCDM) and an $\Omega=1$ standard CDM 
model (SCDM). The clustering amplitude of clusters in OCDM is similar to that
in $\Lambda$CDM, while that in SCDM, although qualitatively similar,
has much lower amplitude, reflecting the relatively small amount of
large-scale power in this model compared to the other three. 

Our predictions may be readily compared with the measured values of
$r_0$ for APM clusters given by Croft et al. (1997). Comparison with
these data is relatively simple because this cluster sample is
approximately volume-limited. By contrast, comparison with X-ray
selected cluster samples (e.g. Ebeling \etal\ 1996, Guzzo \etal\
1999), which are flux-limited, requires more extensive modelling (see
Moscardini \etal\ 2000).  The measured values of $r_0$ for APM
clusters are in good agreement with the predictions of
$\Lambda$CDM. They lie significantly above the $\tau$CDM predictions
for the {\it smallest} values of $d_{\rm c}$. For the $R\ge 1$ Abell
clusters, with $d_c=52h^{-1}$Mpc, Peacock \& West (1992) estimated
$r_0 = 21.1 \pm 1.3$, which is close to the $\Lambda$CDM predictions
-- $18.5h^{-1}$Mpc from the simulation, or $20h^{-1}$Mpc from the
analytic theory -- and also agrees with the APM results on this scale.

\section{Power Spectra for the Cluster Distribution} \label{power}
We have computed the power spectra for the cluster distribution in our 
two simulations. As a comparison observational sample
we take the APM clusters analyzed by Tadros et al.\ (1998). 
The number density in this sample is 
$3.4\times 10^{-5}\,(h^{-1}\,\mbox{Mpc})^{-3}$ which is 
equivalent to $d_{c} = 30.9\,h^{-1}$\,Mpc. At this separation, 
the $\tau$CDM and $\Lambda$CDM simulations contain 
samples of about 270,000 and 915,000 clusters respectively.
The upper panels of figures \ref{fig:power_tcdm} and 
\ref{fig:power_lcdm} show cluster power spectra from 
our simulations at this value of $d_{\rm c}$ (filled circles), the
observational points of Tadros et al.\ (open triangles), and power
spectra for the dark matter (open squares). The bias, defined as the
square root of the ratio of cluster to dark matter power spectrum, is
plotted in the lower panels. The power spectra are quite noisy at the
largest scales because of the small number of modes in the simulated
volume. The peak in the power spectrum is nevertheless quite
clear. The bias is nearly constant over a wide range of scales, and
its value is close to that predicted by the SMT formulae (about
15\% below the prediction of eqn~(3)). The agreement is remarkable 
given the simplicity of the analytic arguments. 
The observed power spectra of Tadros et al.\ (1998) lie
above both models by a factor of about 1.5.  This is a little
surprising since the correlation strength given by Croft et al. (1997)
for the corresponding sample is quite similar to that predicted (see
figure \ref{fig:tpcf_r0}). Of course, our numerical results are in
real space, whereas the APM power spectra are measured in redshift
space.  For these large scales, Kaiser's (1987) expression should be
applicable: 
\begin{equation} {P_s \over P_r} = 1 + 2\beta/3 + \beta^2/5, 
\end{equation} 
where $\beta=\Omega^{0.6}/b$ (see Eke \etal 1996). For $\tau$CDM and
$\Lambda$CDM, the correction factors are respectively 1.22 and 1.15,
less than half the observed offset between models and data. The
remaining differences are not large and may reflect residual
systematics in the observational data analysis. Comparing the
observational points with the simulations it appears premature to
argue that a peak in the observed power spectrum has been detected.

\section{Summary} \label{summary}

We have presented results for the second order clustering statistics of
mass-limited samples of galaxy clusters in our Hubble Volume
simulations. These simulations follow the matter distribution
in by far the largest volumes treated to date, and as a result we are
able to estimate clustering statistics with unprecedented precision.
The two simulations we have studied are a $\tau$CDM universe with
$\Omega=1$ and a $\Lambda$CDM universe with $\Omega=0.3$. Both are
consistent with the fluctuation amplitude measured by COBE and
with the observed abundance of rich clusters at $z=0$. Cluster
correlations are very similar in these two models, although slightly
stronger in the low density case. In both cosmologies, the correlation
length of rich clusters increases from 12 -- 13 $h^{-1}$Mpc for
relatively low mass objects with mean separation 30 $h^{-1}$Mpc to
22 -- 27 $h^{-1}$Mpc for rarer and more massive objects with mean
separation 100 $h^{-1}$Mpc. For both models, the power spectrum
of the cluster distribution is accurately parallel to that of the
dark matter for wavenumbers $k = 0.01$ -- $0.1 h$~Mpc$^{-1}$.

We have compared our results with predictions from the
analytic model of Sheth, Mo \& Tormen (2000). When clustering
strengths are compared as a function of the mean separation of the
cluster sample, there are no adjustable parameters and it is thus
remarkable that we find good agreement in all cases.
Correlation lengths are predicted by the analytic model to better
than 10\%, and the mean bias of the power spectrum is predicted
even more accurately on the scales most relevant for real samples.

We have also compared our results with published data on the APM cluster
sample (Croft et al. 1997; Tadros et al. 1998). The observed trend
of clustering with richness is very similar to those
predicted in our CDM models. The observed correlation lengths 
are consistent with those predicted by our $\Lambda$CDM model at
all richness levels, and are also compatible with our $\tau$CDM
model except perhaps for the poorest systems. The published power 
spectrum for the APM clusters agrees in shape with that predicted 
by the two models, but its amplitude is greater by about 50\%. Only
part of this discrepancy can be attributed to redshift
distortion effects. Since the observed spectrum is based on only 364
clusters, it may be prudent to wait for larger samples before
drawing substantive conclusions from this disagreement.

\section*{Acknowledgements}

CSF acknowledges a PPARC Senior Research Fellowship and a Leverhulme
Research Fellowship. PAT is a PPARC Lecturer Fellow. CSF and SDMW are
grateful for the hospitality of the Institute for Theoretical Physics,
Santa Barbara, where the final text of this paper was written.  We
thank Carlton Baugh and Nelson Padilla for providing code to check our
MW and SMT bias calculations and for useful discussions.

\label{lastpage}
\clearpage
 
\label{lastpage}
\end{document}

%% file: mn_junk.tex
\newif\ifAMStwofonts
\newcommand{\figstart}[1]  {\begin{figure} \psfig{#1}}
\newcommand{\lfigstart}[1] {\begin{figure*} \psfig{#1}}
\newcommand{\figend} {\end{figure}}
\newcommand{\lfigend} {\end{figure*}}

\ifoldfss
  \ifCUPmtlplainloaded \else
    \NewTextAlphabet{textbfit} {cmbxti10} {}
    \NewTextAlphabet{textbfss} {cmssbx10} {}
    \NewMathAlphabet{mathbfit} {cmbxti10} {} % for math mode
    \NewMathAlphabet{mathbfss} {cmssbx10} {} %  "   "    "
  \fi
  \ifAMStwofonts
    \ifCUPmtlplainloaded \else
      \NewSymbolFont{upmath} {eurm10}
      \NewSymbolFont{AMSa} {msam10}
      \NewMathSymbol{\upi}     {0}{upmath}{19}
      \NewMathSymbol{\umu}     {0}{upmath}{16}
      \NewMathSymbol{\upartial}{0}{upmath}{40}
      \NewMathSymbol{\leqslant}{3}{AMSa}{36}
      \NewMathSymbol{\geqslant}{3}{AMSa}{3E}

       \let\ge=\geqslant
    \fi
  \fi
\fi % End of OFSS

\ifnfssone
  \newmathalphabet{\mathit}
  \addtoversion{normal}{\mathit}{cmr}{m}{it}
  \addtoversion{bold}{\mathit}{cmr}{bx}{it}
  \newmathalphabet{\mathbfit} % math mode version of \textbfit{..}
  \addtoversion{normal}{\mathbfit}{cmr}{bx}{it}
  \addtoversion{bold}{\mathbfit}{cmr}{bx}{it}
  \newmathalphabet{\mathbfss} % math mode version of \textbfss{..}
  \addtoversion{normal}{\mathbfss}{cmss}{bx}{n}
  \addtoversion{bold}{\mathbfss}{cmss}{bx}{n}
  \ifAMStwofonts
    \ifCUPmtlplainloaded \else
      \UseAMStwoboldmath
      \makeatletter
      \new@mathgroup\upmath@group
      \define@mathgroup\mv@normal\upmath@group{eur}{m}{n}
      \define@mathgroup\mv@bold\upmath@group{eur}{b}{n}
      \edef\UPM{\hexnumber\upmath@group}
      \new@mathgroup\amsa@group
      \define@mathgroup\mv@normal\amsa@group{msa}{m}{n}
      \define@mathgroup\mv@bold\amsa@group{msa}{m}{n}
      \edef\AMSa{\hexnumber\amsa@group}
      \makeatother
      \mathchardef\upi="0\UPM19
      \mathchardef\umu="0\UPM16
      \mathchardef\upartial="0\UPM40
      \mathchardef\leqslant="3\AMSa36
      \mathchardef\geqslant="3\AMSa3E

       \let\ge=\geqslant
    \fi
  \fi
\fi % End of NFSS release 1

\ifnfsstwo
  \DeclareMathAlphabet{\mathbfit}{OT1}{cmr}{bx}{it}
  \SetMathAlphabet\mathbfit{bold}{OT1}{cmr}{bx}{it}
  \DeclareMathAlphabet{\mathbfss}{OT1}{cmss}{bx}{n}
  \SetMathAlphabet\mathbfss{bold}{OT1}{cmss}{bx}{n}
  \ifAMStwofonts
    \ifCUPmtlplainloaded \else
      \DeclareSymbolFont{UPM}{U}{eur}{m}{n}
      \SetSymbolFont{UPM}{bold}{U}{eur}{b}{n}
      \DeclareSymbolFont{AMSa}{U}{msa}{m}{n}
      \DeclareMathSymbol{\upi}{0}{UPM}{"19}
      \DeclareMathSymbol{\umu}{0}{UPM}{"16}
      \DeclareMathSymbol{\upartial}{0}{UPM}{"40}
      \DeclareMathSymbol{\leqslant}{3}{AMSa}{"36}
      \DeclareMathSymbol{\geqslant}{3}{AMSa}{"3E}

       \let\ge=\geqslant
    \fi
  \fi
\fi % End of NFSS release 2

\ifCUPmtlplainloaded \else
  \ifAMStwofonts \else % If no AMS fonts
    \def\upi{\pi}
    \def\umu{\mu}
    \def\upartial{\partial}
  \fi
\fi

\def\hmpc{h^{-1}{\rm Mpc}}

\def\etal{{et al}}